\documentclass[twoside]{article}
\usepackage{graphicx,amssymb,mathrsfs,amsmath}
\input vatola.sty \input cyracc.def
\input psfig.sty

\usepackage{graphicx}
\usepackage{pdfpages}

\newcommand{\sanhao}{\fontsize{19.08pt}{3\baselineskip}\selectfont}
\newcommand{\xiaosihao}{\fontsize{12pt}{\baselineskip}\selectfont}
\newcommand{\wuhao}{\fontsize{10.5pt}{\baselineskip}\selectfont}
\newcommand{\xiaowuhao}{\fontsize{9.5pt}{\baselineskip}\selectfont}
\newcommand{\dawuhao}{\fontsize{10pt}{0.8\baselineskip}\selectfont}
\newcommand{\liuhao}{\fontsize{7.875pt}{\baselineskip}\selectfont}

\newcommand{\qihao}{\fontsize{7pt}{\baselineskip}\selectfont}
\newcommand{\bahao}{\fontsize{8pt}{\baselineskip}\selectfont}
\TagsOnRight

\renewcommand{\baselinestretch}{1.06} 
\catcode`@=11 \long\def\@makefntext#1{\noindent #1}
\newskip\tabcentering \tabcentering=1000pt plus 1000pt minus 1000pt
\def\REF#1{\par\hangindent\parindent\indent\llap{#1\enspace}}
\def\MCH#1#2{\setbox0=\hbox{\raise#1\hbox{#2}}\smash{\box0}}

\def\@evenfoot{}\def\@oddfoot{}
\def\@evenfoot{\vbox{\hbox to \textwidth{\bahao\sf\hbox to
0.01cm{\textbf{\thepage}\hfill} \hfill{\emph{Li \& Fan Sci China
Ser G-Phys Mech Astron } {$|$ March 2009 $|$ vol. xx $|$ no. xx $|$
\textbf{xxxx-xxxx}} }\hfill}}}
\def\@oddfoot{\vbox{\hbox to \textwidth{\bahao\sf\hbox to
0.01cm{} \hfill{ \emph{\hspace{8mm}Li \& Fan Sci China Ser G-Phys
Mech Astron } {$|$ March 2009 $|$ vol. xx $|$ no. xx $|$
\textbf{xxxx-xxxx}} }\hfill\hfill\textbf{\thepage}}}}

\def\sec#1{\vspace{6mm}\noindent{{\xiaosihao\sf\textbf{#1}}}\vspace{2mm}}

\def\geq{\geqslant}
       
  \def\tlj{\end{document}}  \newsymbol\wjzhml 203F

\def\wj{\end{document}}
\def\no{\noindent}

\begin{document}
\abovedisplayskip=5pt plus 1pt minus 2pt 
\belowdisplayskip=5pt plus 1pt minus 2pt 
\textwidth=145truemm \textheight=212truemm
\renewcommand{\baselinestretch}{0.9}\baselineskip 9pt
{\psfig{figure=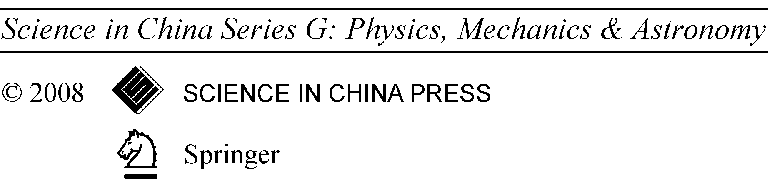}\hfill
\begin{picture}(43,0)
\rightline{\sf\put(-50,25){{\vbox{\hbox {\hspace{5.85mm}\dawuhao
www.scichina.com} \hbox{\hspace{6.5mm}\dawuhao phys.scichina.com}
\hbox{\,\dawuhao www.springerlink.com}}}}}
\end{picture}

\vspace{11.6true mm}
\renewcommand{\baselinestretch}{1.6}\baselineskip 19.08pt
\noindent{\sanhao{\sf\textbf{Limitations of Absolute Current Densities Derived from the Semel \& Skumanich Method}}}
\vspace{0.5 true cm}
\renewcommand{\sfdefault}{phv}

\noindent{\sf  Jing Li$^{\dag}$, Yuhong Fan$^\ddag$
\footnotetext{ \baselineskip 6pt \qihao
\vspace{-2.2mm}\\
Received October 2008; accepted February 2009\\
$^\dag$Corresponding author (email: jingucla@gmail.com) \\
$\ddag$(email: yfan@hao.ucar.edu)
\vspace{2mm}\\

\centerline{\bahao\sf \emph{Sci China Ser G-Phys Mech Astron} $|$
March 2009 $|$ xxx $|$ xxx $|$ \textbf{xxx}}
}}

\vspace{0.2 true cm}\noindent
\parbox{13.3cm}
{\noindent\renewcommand{\baselinestretch}{1.3}\baselineskip 12pt
{\liuhao\sf
Institute for Astronomy, University of Hawaii, 2680 Woodlawn Drive, Honolulu, HI 96822 USA\\ 
High Altitude Observatory, Earth and Sun Systems Laboratory, National Center for Atmospheric Research, P.O. Box 3000, Boulder, CO 80307 USA\vspace{2mm}}}

\noindent{\xiaowuhao\sf\textbf{\hspace{-1mm}
\parbox{13.3cm}
{\noindent
\renewcommand{\baselinestretch}{1.3}\baselineskip 13pt
Semel and Skumanich$^{[1]}$ proposed a method to obtain the absolute electric current density, $|J_z|$, without disambiguation of $180^\circ$ in the transverse field directions. The advantage of the method is that the uncertainty in the determination of the ambiguity in the magnetic azimuth is removed. Here, we investigate the limits of the calculation when applied to a numerical MHD model$^{[2,3]}$. We found that the combination of changes in the magnetic azimuth with vanishing horizontal field component leads to errors, where electric current densities are often strong. Where errors occur,  the calculation gives $|J_z|$ too small by factors typically $1.2 \sim 2.0$.}}}

\vspace{5.5mm}\no{\footnotesize \sf
Numerical MHD Model, Magnetic Field, Electric Current Density} \vspace{6mm}
\baselineskip 15pt

\renewcommand{\baselinestretch}{1.08}
\parindent=10.8pt  
\rm\wuhao\vspace{-4mm}

\sec{1\quad Introduction}

\noindent The electric current density is an important measure of the non-potentiality of the magnetic field in solar active regions. It has the potential to illuminate the active region eruption process. Vertical electric current densities can be calculated from vector magnetic fields measured in the photosphere.  They have been repeatedly obtained ever since photospheric vector magnetic fields first became routinely available in the 1980s$^{[4,5,6,7,8,9,10,11,12,13,14,15,16,17,18]}$. All these calculations involve the resolution of the $180^\circ$ ambiguity in the transverse field directions, which problem is caused by the diagnosis of the magnetic field via the Zeeman effect.  Semel \& Skumanich (1998)$^{[1]}$ developed a method to calculate the absolute vertical electric current density from observed vector magnetic fields. Their formulation is especially interesting in that it does not require the $180^\circ$ disambiguation of the  transverse field directions and therefore removes one uncertainty in the current calculation. 

The Semel \& Skumanich method can be outlined as follows. In SI units, Amp\`{e}re's law reads $\mu_0 {\bf J} = \nabla \times {\bf B}$, where ${\bf J}$ is the current density and $\mu_0$ is the permittivity of the vacuum. Since $\mu_0 = 4 \pi \times 10^{-7}$ [T m A$^{-1}$], then $4 \pi J_z =\frac{\partial B_y} {\partial x} - \frac{\partial B_x} {\partial y}$, where the magnetic field is measured in G and the current density is measured in mA m$^{-2}$, and the distances are measured in m. Multiplying both sides of the $J_z$ equation by $B_x^4$, $B_y^4$, and $B_xB_y$, $B_yB_x$, an expression for $[(\nabla\times {\bf B})_z]^2$ is obtained
\begin{equation}
[(\nabla\times {\bf B})_z]^2=B_x^2g_y^2+B_y^2g_x^2-B_\bot^2g_xg_y\sin2\phi
\label{eq:jz2}
\end{equation}
which is equivalent to
\begin{equation}
|\nabla\times {\bf B}|_z = B_xg_y-B_yg_x\\
\end{equation}
Here, $g_x,g_y$ are given by
\begin{eqnarray}
B_\bot^4g_x=B_y^2\frac{\partial(B_xB_y)}{\partial y}-\frac{1}{2}B_xB_y\frac{\partial(B_y^2-B_x^2)}{\partial y}-\frac{1}{2}B_\bot^2\frac{\partial B_y^2}{\partial x} \\
B_\bot^4g_y=B_x^2\frac{\partial(B_xB_y)}{\partial x}-\frac{1}{2}B_xB_y\frac{\partial(B_x^2-B_y^2)}{\partial x}-\frac{1}{2}B_\bot^2\frac{\partial B_x^2}{\partial y}
\end{eqnarray}
\noindent where $B_\bot=\sqrt{B_x^2+B_y^2}$, the strength of the horizontal magnetic component; $B_x=B_\bot\cos\phi$ and $B_y=B_\bot\sin\phi$ are two perpendicular horizontal components and $\phi$ is the magnetic azimuth. All are observable quantities, but $[(\nabla\times {\bf B})_z]^2$ does not vary with either $\phi$ or $\phi+180^\circ$. 

Recently$^{[19]}$, we used Equation (2) to calculate the absolute vertical current densities in two morphologically different active regions: a simple sunspot region free of flare activity (NOAA 10001 on 20 June 2002); and an active region of the quadrupolar configuration with multiple flares (including a white light X3 flare$^{[20]}$) and coronal mass ejections (NOAA 10030 on 15 July 2002). We found that $|J_z|$ correlated with vertical magnetic field, $B_\parallel$, in the form $|J_z|=a+b |B_\parallel|$ for both regions on large scales.  The relationship between $|J_z|$ and $|B_\parallel|$ needs to be examined with more samples of active regions, and observational data from different instruments. Nevertheless, we realize that the use of $|J_z|$ has the potential to characterize active regions in their flare productivity.  The absolute vertical current density is particularly useful for studying the mechanical forces due to currents induced in moving material$^{[21]}$. 


Equation (2) is limited by the lack of disambiguation of the $180^\circ$ in the magnetic azimuth. Semel \& Skumanich$^{[1]}$ pointed out that the calculation ``blindly'' follows the magnetic field continuity, leading to failure when (1) the magnetic azimuth suddenly changes direction; (2) the horizontal component simply vanishes paralyzing the continuous magnetic field assumption.  In this short paper, we explore the limits of the $|J_z|$ calculation with reference to a MHD numerical simulation developed by Fan \& Gibson$^{[2,3]}$. We will describe the MHD numerical simulation in the next section. In section 3, we will present errors of Equation (2), and show statistics over pixels. A discussion is given in section 4, and a summary in section 5.

\begin{center}
\begin{figure}
\centering
\includegraphics[width=1.0\textwidth]{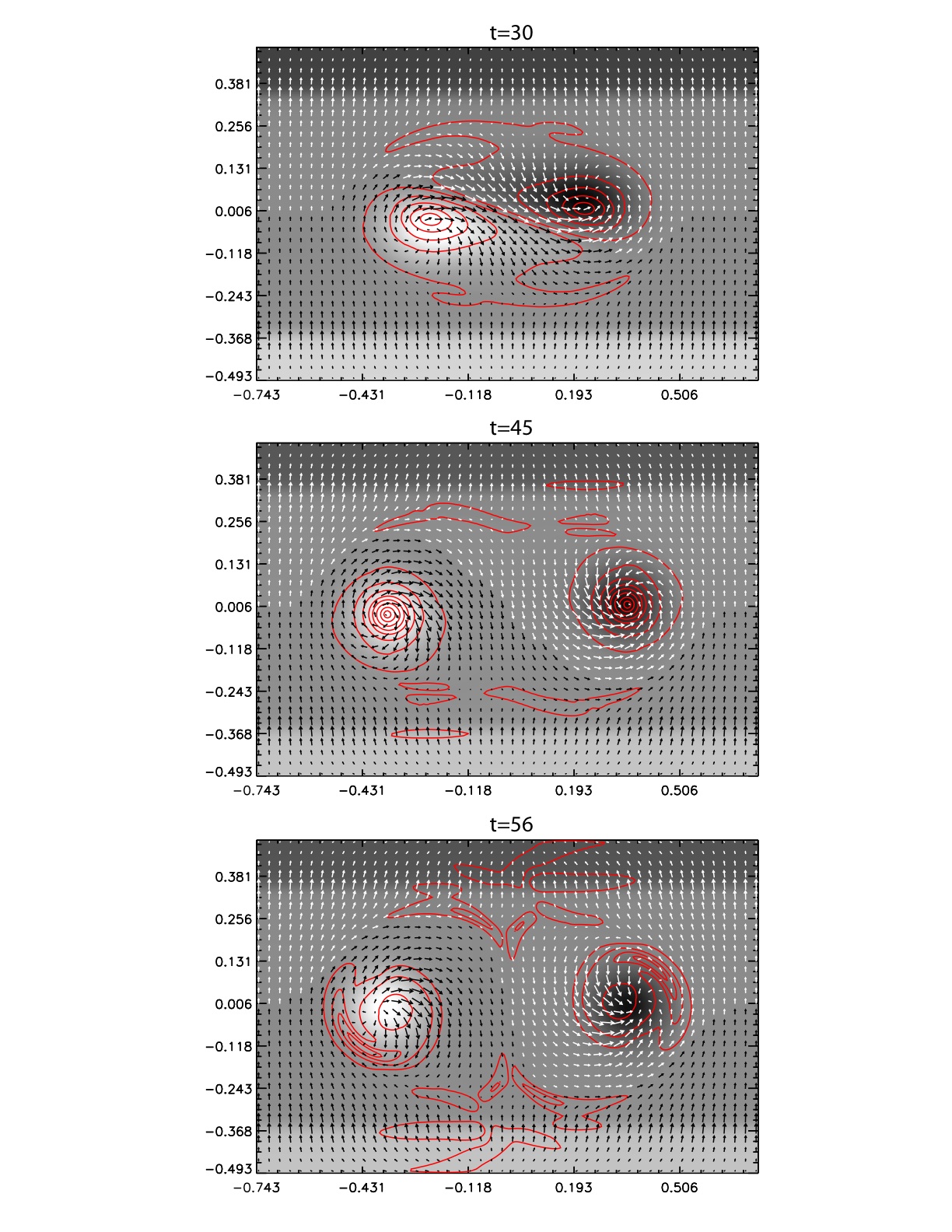}
\caption{\footnotesize \quad The MHD simulations at $t=30,45,56$ in the xy-plane above the lower boundary, $z=0.00625L$. The background images are the vertical magnetic field, $B_z$. The horizontal magnetic fields are plotted in arrows: white arrows correspond to $B_z<0$, and black arrows correspond to $B_z\geq 0$. Directions of arrows represent ${\bf B}_\bot$ directions, and the lengths of the arrows represent ${\bf B}_\bot$ strengths.  The absolute vertical current densities, $|J_z|$, are plotted in contours, $|J_z|=0.01,0.04,0.08,0.12,0.15,0.18,0.21,0.23$.\mbox{}\\ \mbox{}}
\end{figure}
\end{center}

\sec{2\quad MHD Model}

\noindent A 3-dimensional numerical simulation of the coronal response to a rising flux tube was presented by Fan \& Gibson$^{[2,3]}$. The system reaches instability when the flux tube contains a twist of $1.875\times 2\pi$ turns about the tube axis between two footpoints when fully emerged. In this work, we use the simulation at three times corresponding to different phases of the flux tube evolution. At $t=30$ (all times are given in arbitrary units), the flux tube axis rises above the lower boundary at a sub-Alfv\'{e}nic speed. At $t=45$, the flux tube contains $1.5\times 2\pi$ turns between the footpoints above the lower boundary. The tube undergoes a significant acceleration with both writhing and rising motions, but it stays in a 2-dimensional plane. At $t=56$, the flux tube contains $1.875\times 2\pi$ twists between the two footpoints. The simulation reached the critical point at the onset of the kink instability. The complexity of the model is represented by the turns of the flux tube, which grows with increasing time. 

The model is built in a Cartesian domain in a box with resolution, $240\times160\times200$ in the units $x=[-0.75L,0.75L],y=[-0.5L,0.5L],z=[0,1.25L]$, where $L$ is the length of the box edges. In our work, the ``photospheric magnetic field'' is the field at $z=0.003125L$ just above the lower boundary. Figure 1 shows the 2-dimensional magnetic fields in the model at the three times. The vertical magnetic fields, $B_z$, are the background images. The horizontal fields, ${\bf B}_\bot$, are plotted with short arrow bars: white arrow bars correspond to $B_z<0$,  and black ones correspond to $B_z\geq0$. $|J_z|$ are plotted as contours over the magnetic fields. 

\sec{3\quad Errors in the $|J_z|$ Calculation}

\noindent  Errors from the application of Equation (2) are evaluated by comparing the true ($|J_z|$) with the derived current density  ($|J_z|_{ss}$) after excluding the current density on the edges of the simulation box (it is apparent that the current densities cannot be calculated on edges of the MHD simulation box where $B_x(x,y)$ and $B_y(x,y)$ are discontinued). $|J_z|$ is calculated from $4 \pi J_z =\frac{\partial B_y} {\partial x} - \frac{\partial B_x} {\partial y}$, taking the absolute value.  $|J_z|_{ss}$ is calculated with Equation (2), where ``$ss$'' represents authors of the method, Semel \& Skumanich. In this work, the derivatives are replaced with the differences between neighboring pixels on the x-y plane: 
\begin{equation}
         \frac{\partial B_y}{\partial x}-\frac{\partial B_x}{\partial y}\approx\frac{B_y(x+\Delta x,y)-B_y(x-\Delta x,y)}{2\Delta x}-\frac{B_x(x,y+\Delta y)-B_x(x,y-\Delta y)}{2\Delta y}
\end{equation}

\noindent where $\Delta x$ and $\Delta y$ are the grid unit on the x- and y-axes, respectively. The approximation of derivatives to differences gives rise to numerical errors. These errors are introduced to both true $|J_z|$ and derived current densities $|J_z|_{ss}$; but they are squared in the $|J_z|_{ss}$ because the differences are carried out for $B_x^2, B_y^2, B_xB_y$ between neighboring pixels in Equation (2). The smaller are the $\Delta x$ and $\Delta y$, the better are the approximations. 

\begin{center}
\begin{figure}
\centering
\includegraphics[width=1.0\textwidth]{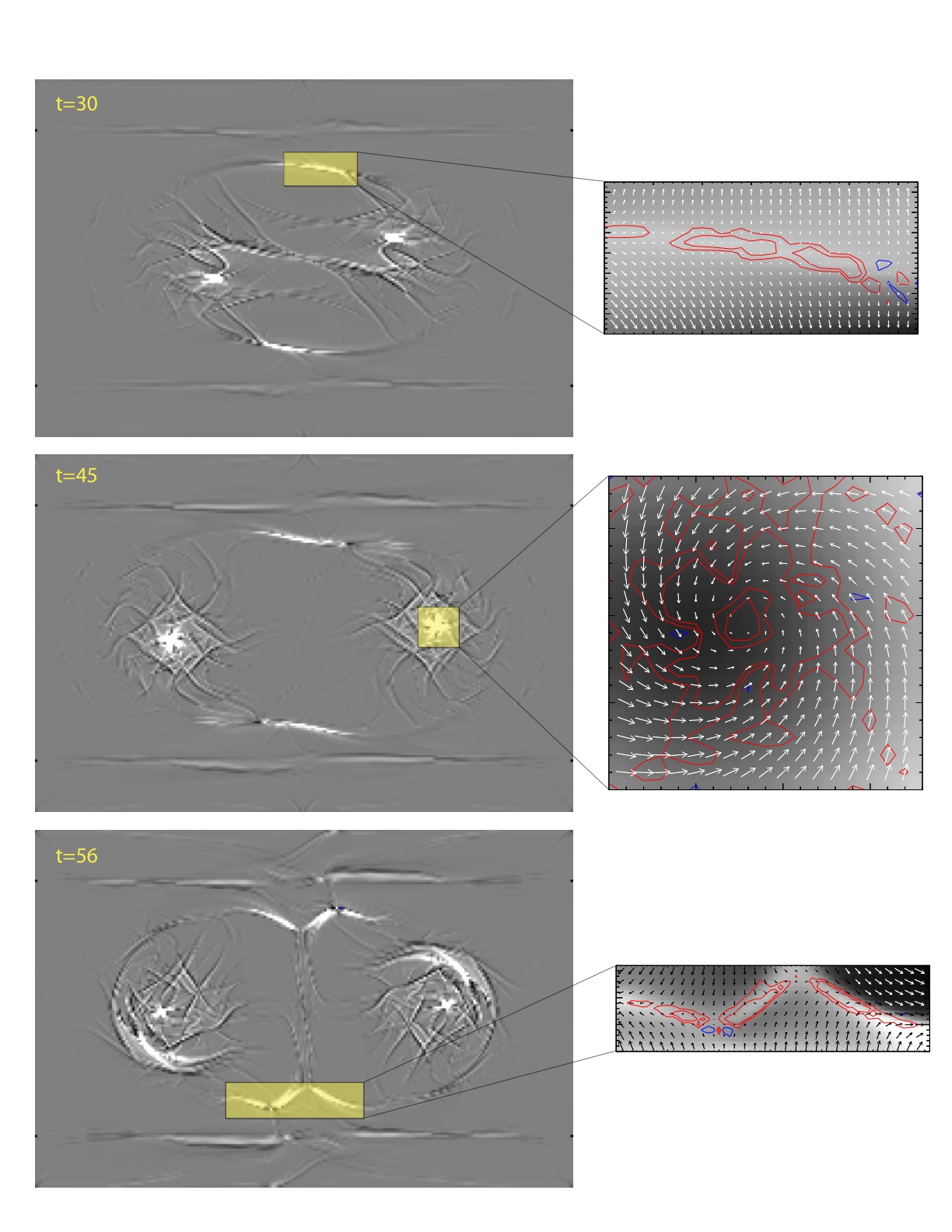}
\caption{\footnotesize \quad Demonstration of Type 1 error. Left: $\Delta J_z=|J_z|-|J_z|_{ss}$ at three times $T=30,45,56$ from the top to the bottom; The black- and white-most brightness represent [-0.005,0.005]. Small yellow boxes highlight the area where $|J_z|$ and $|J_z|_{ss}$ disagree with large uncertainties. Right: Vector magnetic fields within corresponding highlighted yellow boxes, which bars and the background images have the same meanings as those in the Fig.1. The contours represent $\Delta J_z=|J_z|-|J_z|_{ss}=-0.1,-0.05,-0.005,-0.002,0.002,0.005,0.05,0.1$. Red contours represent  $\Delta J_z>0$ and blue contours represent $\Delta J_z<0$.\mbox{}\\ \mbox{}}
\end{figure}
\end{center}

Figure 2 shows maps of $\Delta J_z= |J_z|-|J_z|_{ss}$. These maps highlight pixels where large disagreements between $|J_z|$ and $|J_z|_{ss}$ occur. Figure 3 shows maps of $\Delta J_z/J_z$ representing the fractional errors in derived current density. Small values of $|\Delta J_z|$ over weak current densities are given equal erroneous impressions to those large values of $|\Delta J_z|$ over strong current densities. 

\begin{center}
\begin{figure}
\centering
\includegraphics[width=1.0\textwidth]{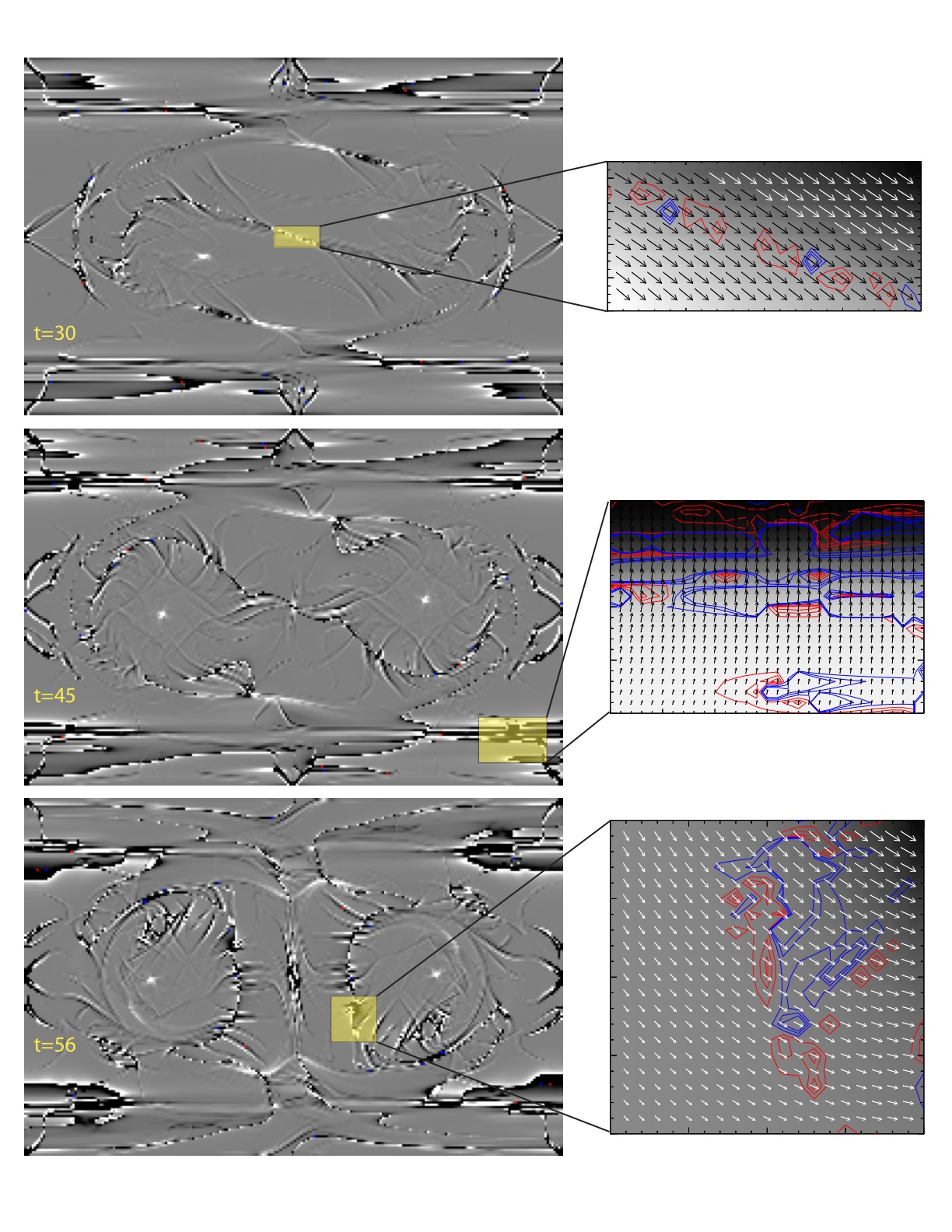}
\caption{\footnotesize \quad Demonstration of Type 2 error. Left: $\Delta J_z/J_z$ at three times $T=30,45,56$ from the top to the bottom; The black- and white-most brightness represent [-0.5,0.5]. Small yellow boxes highlight the area where large ratios $\Delta J_z/J_z$ occur. Right: Vector magnetic fields within corresponding highlighted yellow boxes: background image is the vertical magnetic components; the short bars represent the horizontal magnetic fields. The contours represent $\Delta J_z=|J_z|-|J_z|_{ss}=-0.75,-0.5,-0.2,0.2,0.5,0.75$. Red contours represent  $\Delta J_z/J_z>0$ and blue contours represent $\Delta J_z/Jz<0$.\mbox{}\\ \mbox{}}
\end{figure}
\end{center}

Table 1 presents quantitative evaluations of Equation (2) when the current densities are calculated with models at the times $t=30,45,56$. The analyses are conducted over pixels selected by two criteria: $|\Delta J_z|\geq0.005$ (Case 1) and  $|\Delta J_z/J_z|\geq0.5$ (Case 2).  The percentages of problematic pixels in either case are listed in columns 1st and 5th of Table 1. The average errors on $|J_z|_{ss}$ calculation are calculated by taking median values of $|\Delta J_z|$ over selected pixels for cases 1 and 2 (2nd and 6th columns, respectively). The medians of [$|J_z|,|J_z|_{ss}$]  are presented in the 3rd, 4th, and 7th, 8th columns, respectively, over respective sets of pixels defined by either $|\Delta J_z|\geq0.005$ or $|\Delta J_z/J_z|\geq0.5$.  Figure 4 shows the correlation between $|J_z|$ (x-axis) and $|J_z|_{ss}$ (y-axis). All pixels are represented by a dot. Pixels where $|\Delta J_z|\geq 0.005$ are marked by ``$\Diamond$'' symbols as well. 


\noindent{{\bf Table 1}\quad Statistical Analyses over Pixels }\\
{\footnotesize \vspace{1.5mm} \doublerulesep 0.4pt \tabcolsep 5.5pt
\begin{tabular*}{\textwidth}{ccccccccc}
\hline \hline
& 
\multicolumn{4}{c}{Case 1: Pixels ($|\Delta J_z|\geq0.005$)} &
\multicolumn{4}{c}{Case 2: Pixels ($|\Delta J_z/J_z|\geq0.5$)} \\
&
\multicolumn{4}{c}{------------------------------------------------------} &
\multicolumn{4}{c}{-----------------------------------------------------------------------} \\
& &
\multicolumn{3}{c}{{\it median}} &
&
\multicolumn{3}{c}{{\it median}} \\
 Time  & 
 Pixel \% & $|\Delta J_z|$ &$|J_z|$  & $|J_z|_{ss}$& 
 Pixel \% & $|\Delta J_z|$ &$|J_z|$  &$|J_z|_{ss}$\\\hline
  30  &  0.308 & $8.97\times 10^{-3}$ & 0.092 & 0.049&3.941 &$5.54\times 10^{-5}$&$5.09\times 10^{-5}$&$6.13\times 10^{-5}$\\
 45  &  0.425 & $8.35\times10^{-3}$ & 0.193 & 0.166 & 6.515&$8.42\times 10^{-5}$&$5.70\times 10^{-5}$&$9.42\times 10^{-5}$\\
 56  &  0.675 & $7.97\times 10^{-3}$ & 0.060 & 0.047 & 7.143&$1.22\times 10^{-4}$&$8.05 \times 10^{-5}$&$1.22\times 10^{-4}$\\
 column& 1st&2nd&3rd&4th&5th&6th&7th&8th\\\hline
 \hline \hline
\end{tabular*}}

\begin{center}
\begin{figure}
\centering
\includegraphics[width=0.6\textwidth]{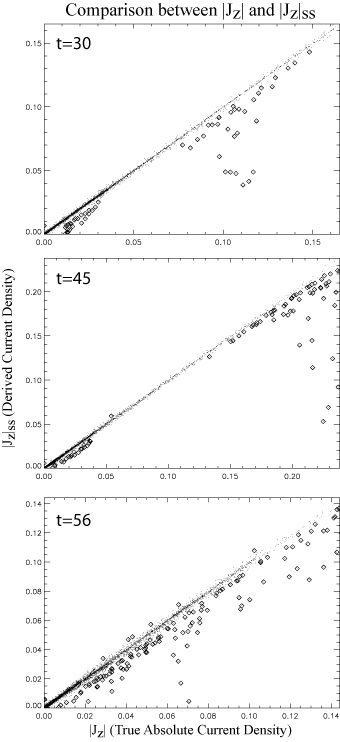}
\caption{\footnotesize \quad $|J_z|$ versus $|J_z|_{ss}$ at three times, $t=30,45,56$. The dots represent all pixels within the x-y plane. The ``$\Diamond$'' symbols represent pixels having $|\Delta J_z|>0.005$, most of them represent Type 1 errors.\mbox{}\\ \mbox{}}
\end{figure}
\end{center}

\sec{4\quad Discussion}

\noindent Within entire simulation planes, we found that the average $|J_z|_{ss}$ and $|J_z|$ are very close to each other. The {\it r.m.s.}($|J_z|$) are 0.02192, 0.0251 and 0.0211 for models at t=30, 45 and 56, respectively. In comparison, the {\it r.m.s.}($|J_z|_{ss}$) are 0.0168, 0.0244, and 0.0207.  They are slightly smaller than those $|J_z|$ by a factor typically $1.01\sim1.03$. On the other hand, large differences between $|J_z|$ and $|J_z|_{ss}$ occur on some special pixels which are demonstrated in Cases 1 and 2.

Numbers of pixels in Case 1 are only a few per thousand pixels (1st column in Table 1). Over these pixels, the median $|J_z|$ (3rd column) are stronger than average $|J_z|$ of the entire simulation planes by factors $12, 24$, and $7$ suggesting strong current densities concentrate on these pixels.  Examination of Table 1 and Figs. 1 and 2 shows that the combination of vanishing horizontal field strength and dramatic changes in the magnetic azimuth results in large errors. Examples of problematic pixels are marked by yellow rectangles in Fig. 2.  On the right hand side, the rectangles are enlarged to demonstrate the horizontal magnetic field in all pixels. The vertical magnetic components are the background images, and the $\Delta J_z$ are plotted in contours. This confirms what had been discussed by Semel \& Skumanich$^{[1]}$ that the method fails on pixels where the magnetic field lines are discontinuous. We call errors originated from discontinuous field lines Type 1 error. 

Comparing the 3rd with the 4th columns in Table 1, the Type 1 error results in that $|J_z|_{ss}$ underestimates the true $|J_z|$ by factors $1.9, 1.2$, and $1.3$. This is also evident in Fig. 4 where the majority of ``$\Diamond$'' symbols fall below the $|J_z|=|J_z|_{ss}$ line.  This can be understood as a result of the ``blind'' calculation of $|J_z|_{ss}$. The dramatic change in the magnetic azimuth causes increasing magnitude in $|J_z|$, but Equations (1) and (2) assume the most close magnetic azimuths between neighboring pixels, resulting in smaller $|J_z|_{ss}$. 

Numbers of pixels in Case 2 are a few percent of the total number of pixels (5th column in Table 1). Unlike magnetic fields on pixels in Case 1,  the horizontal magnetic fields are neither vanishing nor experiencing dramatic changes in the azimuth on pixels in Case 2. The sample pixels are demonstrated on the right hand side of Fig. 3 corresponding to the yellow rectangles on the $\Delta J_z/J_z$ maps. The median $|J_z|$ (7th column) are only a few hundredth of the average $|J_z|$ indicating that the vertical current densities are generally weak on these pixels. Finite differences are squared in the Equation (2) and are probably responsible for this kind of error. We call it Type 2 error. 

The problematic pixels demonstrated in Case 2 are 10 times more numerous than those pixels demonstrated in Case 1 (1st and 5th columns). But the median differences between $|J_z|$ and $|J_z|_{ss}$ in Case 2 are more than 100 times as small as those in Case 1 (2nd and 6th columns). 
These mean that the dominant errors of the Equation (2) is Type 1 error. In addition, the median $|J_z|$ in Case 1 are much stronger than those median $|J_z|$ in Case 2 by factors 1807, 3385 and 745 (3rd and 7th columns). This means that Type 1 error often occurs  over strong vertical current densities. Type 2 error only becomes prominent where current densities are weak. In real observations, Type 2 error can be omitted because it is 2 magnitudes smaller than that of the Type 1 error.

\sec{5\quad Summary}

\noindent Equations (1) and (2), proposed by Semel \& Skumanich$^{[1]}$, provide one more tool to estimate the vertical current density from vector magnetic field observations in the observing plane. Their method is free of the disambiguation of $180^\circ$ in the transverse field directions, therefore, removes an uncertainty from the $J_z$ calculation. We apply the Semel \& Skumanich formula to a MHD numerical simulation$^{[2,3]}$ to explore the limits of the method. We identified two types of errors to the method. Dramatic changes in the magnetic azimuths with vanishing horizontal fields result in Type 1 error. The approximation of the derivatives to differences results in Type 2 error. A summary is given below.

\begin{enumerate}
\item Equation (2) accurately measures the vertical current density in a majority of pixels ($>95$\%) in the observing plane.

\item Type 1 error results from the combination of vanishing horizontal fields and dramatic changes in the magnetic azimuth. The errors occur because of an inherit insufficiency in the absolute current density calculation proposed by Semel \& Skumanich$^{[1]}$. 

\item Type 2 error results in the approximations of derivatives to differences, which are squared in the Semel \& Skumanich$^{[1]}$ calculations. Errors become prominent where current densities are weak. 

\item Type 1 error is about 100 times as large as Type 2 error. In real observations, Type 2 error can be omitted.


\item When errors occur, the derived current density is smaller than the actual current density by factors typically $1.2\sim 1.9$.

\item The number of unreliable pixels increases with the increasing number of twists in the magnetic flux tube system above the boundary.

\end{enumerate}

\sec{6\quad Acknowledgment}

\noindent JL would like to thank David Jewitt for his comments on the manuscript, and patient editions on the English writing. JL would also like to thank Dr. Yu Lu for his comments which greatly improved the paper. University Research Council (URC), University of Hawaii, supported JL's travel to attend the Solar Eclipse Conference in Jiuquan, China, July-August 2008.

\normalsize \vskip0.16in\parskip=0mm \baselineskip 15pt
\renewcommand{\baselinestretch}{1.12}

\footnotesize
\parindent=6mm

\bahao\REF{1\  }Semel, M., \& Skumanich, A.\ 1998,A\&A, 331, 383

\REF{2\ }Fan, Y., and Gibson, S. E., 2003, ApJL, 589, L105

\REF{3\ }Fan, Y., and Gibson, S.E., 2004, ApJ, 609, 1123

\REF{4\ }Deloach, A.~C., Hagyard, M.~J., Rabin, D., Moore, R.~L., Smith, B.~J., Jr., West, E.~A., \& Tandberg-Hanssen, E.\ 1984, Sol. Phys., 91, 235 

\REF{5\ }Lin, Y., \& Gaizauskas, V.\ 1987, Sol. Phys., 109, 81

\REF{6\ }Hagyard, M.~J.\ 1988, Sol. Phys., 115, 107 

\REF{7\ }Canfield, R.~C., et  al.\ 1992, PASJ, 44, L111 

\REF{8\ }Leka, K.~D., Canfield, R.~C., McClymont, A.~N., de La Beaujardiere, J.-F., Fan, Y., \& Tang, F.\ 1993, ApJ, 411, 370 

\REF{9\ }de La Beaujardiere, J.-F., Canfield, R.~C., \& Leka, K.~D.\ 1993, ApJ, 411, 378 

\REF{10\ }Zhang, H., \& Wang, T.\ 1994, Sol. Phys., 151, 129

\REF{11\ }Wu, S.T., Weng, F.S., Wang, H.M., Ziron, H., and Ai, G.X., 1993, Advance in Space Res., 13, 127

\REF{12\ }van Driel-Gesztelyi, L., Hofmann, A., Demoulin, P., Schmieder, B., \& Csepura, G.\ 1994, Sol. Phys., 149, 309 

\REF{13\ }Metcalf, T.~R., Canfield, R.~C., Hudson, H.~S., Mickey, D.~L., Wulser, J.-P., Martens, P.~C.~H., \& Tsuneta, S.\ 1994, ApJ, 428, 860 

\REF{14\ }Gary, G.~A., \& Demoulin, P.\ 1995, ApJ, 445, 982 

\REF{15\ }Li, J., Metcalf, T.~R.,  Canfield, R.~C., Wuelser, J.-P., \& Kosugi, T.\ 1997, ApJ, 482, 490 

\REF{16\ }Burnette, A.~B., Canfield, R.~C., \& Pevtsov, A.~A.\ 2004, ApJ, 606, 565 

\REF{17\ }Gary, A., and Moore, R.L.,2004, ApJ, 611, 545

\REF{18\ }Gao, Y., Xu, H., \& Zhang, H.\ 2008, Advances in Space Research, 42, 888 

\REF{19\ }Li, J., van Ballegooijen, A., and Mickey, D., 2009, ApJ, 692, 1543

\REF{20\ }Li, J., Mickey, D.~L., \& LaBonte, B.~J.\ 2005, ApJ, 620, 1092 

\REF{21\ }Cowling, T.~G.\ 1945, Royal Society of London Proceedings Series A, 183, 453






 \tlj

\begin{center}
\centerline{\psfig{figure=fig1.eps}}
\centerline{\footnotesize Fig. 1.\quad } 
\end{center}

\parbox[c]{60mm}{\centerline{\psfig{figure=fig1.eps}}
\centerline{\footnotesize Fig. 1.\quad The MHD simulations at $t=30,45,56$ in the xy-plane above the lower boundary, $z=0.00625L$. The background images are the vertical magnetic field, $B_z$. The horizontal magnetic fields are plotted in arrows: white arrows correspond to $B_z<0$, and black arrows correspond to $B_z\geq 0$. Directions of arrows represent ${\bf B}_\bot$ directions, and the lengths of the arrows represent ${\bf B}_\bot$ strengths.  The absolute vertical current densities, $|J_z|$, are plotted in contours, $|J_z|=0.01,0.04,0.08,0.12,0.15,0.18,0.21,0.23$.}}
\parbox[c]{60mm}

\begin{center}
\centerline{\psfig{figure=fig2.eps}}
\centerline{\footnotesize Fig. 2.\quad Demonstration of Type 1 error. Left: $\Delta J_z=|J_z|-|J_z|_{ss}$ at three times $T=30,45,56$ from the top to the bottom; The black- and white-most brightness represent [-0.005,0.005]. Small yellow boxes highlight the area where $|J_z|$ and $|J_z|_{ss}$ disagree with large uncertainties. Right: Vector magnetic fields within corresponding highlighted yellow boxes, which bars and the background images have the same meanings as those in the Fig.1. The contours represent $\Delta J_z=|J_z|-|J_z|_{ss}=-0.1,-0.05,-0.005,-0.002,0.002,0.005,0.05,0.1$. Red contours represent  $\Delta J_z>0$ and blue contours represent $\Delta J_z<0$.} 
\end{center}

\begin{center}
\centerline{\psfig{figure=fig3.eps}}
\centerline{\footnotesize Fig. 3.\quad Demonstration of Type 2 error. Left: $\Delta J_z/J_z$ at three times $T=30,45,56$ from the top to the bottom; The black- and white-most brightness represent [-0.5,0.5]. Small yellow boxes highlight the area where large ratios $\Delta J_z/J_z$ occur. Right: Vector magnetic fields within corresponding highlighted yellow boxes: background image is the vertical magnetic components; the short bars represent the horizontal magnetic fields. The contours represent $\Delta J_z=|J_z|-|J_z|_{ss}=-0.75,-0.5,-0.2,0.2,0.5,0.75$. Red contours represent  $\Delta J_z/J_z>0$ and blue contours represent $\Delta J_z/Jz<0$.} 
\end{center}

\begin{center}
\centerline{\psfig{figure=fig4.eps}}
\centerline{\footnotesize Fig.4.\quad $|J_z|$ versus $|J_z|_{ss}$ at three times, $t=30,45,56$. The dots represent all pixels within the x-y plane. The ``$\Diamond$'' symbols represent pixels having $|\Delta J_z|>0.005$, most of them represent Type 1 errors.} 
\end{center}

\bf {\boldmath